\documentclass[aps,amsmath,amssymb,prc,preprint,superscriptaddress,showkeys]{revtex4-1}

\usepackage{graphicx}
\usepackage{dcolumn}
\usepackage{bm}
\usepackage{siunitx}
\newcommand{\nuc}[2]{\hbox{$^{#1}$#2}}
\usepackage{gensymb}
\DeclareSIUnit[number-unit-product = {}]{\inchQ}{\textquotedbl}
\DeclareSIUnit[number-unit-product = {\thinspace}]{\inch}{in}
\usepackage[dvipsnames]{xcolor}
\usepackage{hyperref}

\begin{document}

\title{Nuclear structure and direct reaction studies in particle-$\gamma$ coincidence experiments at the FSU John D. Fox Superconducting Linear Accelerator Laboratory}

\author{M. Spieker}
 \email{mspieker@fsu.edu}
\affiliation{Department of Physics, Florida State University, Tallahassee, Florida 32306, USA}

 \author{S. Almaraz-Calderon}
  \email{salamarazcalderon@fsu.edu}
\affiliation{Department of Physics, Florida State University, Tallahassee, Florida 32306, USA}

\begin{abstract}

Since its foundation in the 1960s, the John D. Fox Superconducting Linear Accelerator Laboratory at Florida State University (FSU) pursued research at the forefront of nuclear science. In this contribution, we present recent highlights from nuclear structure and reaction studies conducted at the John D. Fox Superconducting Linear Accelerator Laboratory, also featuring the general experimental capabilities at the laboratory for particle-$\gamma$ coincidence experiments. Specifically, we focus on light-ion induced reactions measured with the Super-Enge Split-Pole Spectrograph (SE-SPS) and the CATRiNA neutron detectors, respectively. Some results obtained with the CeBrA demonstrator for particle-$\gamma$ coincidence experiments at the SE-SPS are presented. A highlight from the first experimental campaigns with the combined CLARION2-TRINITY setup, showing that weak reaction channels can be selected, is discussed as well.

\end{abstract}


\maketitle


\section{Introduction}

Nuclear physics has entered a new exciting era with next-generation rare isotope beam facilities like the Facility for Rare Isotope Beams (FRIB) coming online and enabling experiments with atomic nuclei, which were previously inaccessible, to study their structure and a multitude of reactions with them. These experiments are expected to inform, {\it e.g.}, $r$-process nucleosynthesis and to test fundamental symmetries by using nuclei as laboratories enhancing signals to investigate beyond standard model physics. In this new era, stable-beam facilities continue to play an important role by allowing detailed, high-statistics experiments with modern spectroscopy setups and provide complementary information for rare-isotope studies by, {\it e.g.}, studying structure phenomena of stable nuclei close to the particle-emission thresholds and by investigating details of different nuclear reactions, thus, testing reaction theory. Modern coincidence experiments, that combine multiple detector systems, can also address open questions in stable nuclei providing important pieces to solving the nuclear many-body problem and quality data to guide the development of {\it ab-initio}-type theories for the spectroscopy of atomic nuclei.

Since its foundation in the 1960s, the John D. Fox Superconducting Linear Accelerator Laboratory at Florida State University \cite{fox_web} has continued to pursue research at the forefront of nuclear science. New experimental setups, which were recently commissioned at the Fox Laboratory and which will be presented in this article, enable detailed studies of atomic nuclei close to the valley of $\beta$ stability through modern spectroscopy experiments that detect particles and $\gamma$ rays in coincidence.

\subsection{History of the John D. Fox Laboratory}

The Florida State University (FSU) Accelerator Laboratory began operation in 1960 following the installation of an EN Tandem Van de Graaf accelerator. It was the second of its type in the United States. Since its dedication in March 1960, the FSU Accelerator Laboratory has been recognized for several scientific and technical achievements. Examples of the early days of operation are the first useful acceleration of negatively-charged helium ions at FSU in 1961 \cite{Car63a} and the experimental identification of isobaric analogue resonances in proton-induced reactions in 1963 \cite{Fox64a}.


The laboratory entered its second development stage in 1970 with the installation of a Super-FN Tandem Van de Graaff accelerator. As a third major stage of evolution, a superconducting linear post-accelerator based on \textsc{atlas} technology was funded by the U.S. National Science Foundation in the mid-1980s \cite{Cha81a}, with the first experiment on the completed facility run in 1987 \citep{Fox87a, Mye89a}. The Super-FN Tandem Van de Graaff and superconducting linear post-accelerator are still being used at the FSU Accelerator Laboratory today. In combination with two SNICS sources and an RF-discharge source, they provide a variety of accelerated beams, ranging from protons to accelerated titanium ions, for experiments relevant for nuclear science. In March 2007, FSU's Superconducting Linear Accelerator Laboratory was named for John D. Fox, a longtime FSU faculty member who was instrumental in its development. 

\begin{figure}[t]
\begin{center}
\includegraphics[width=.5\linewidth]{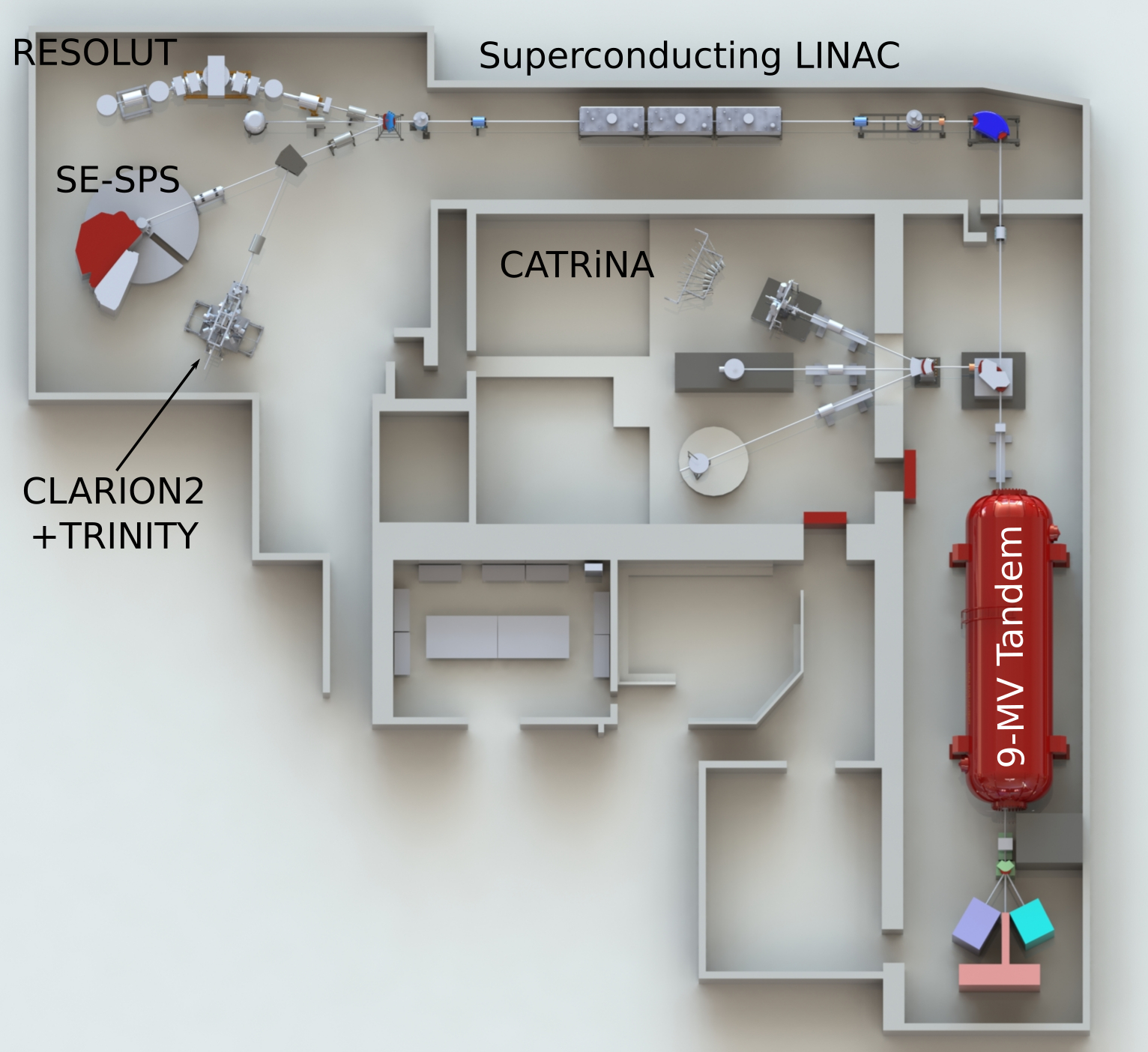}
\end{center}
\caption{Model of the FSU John D. Fox Superconducting Linear Accelerator Laboratory as of fall 2024. Experimental setups featured in this article are highlighted.}\label{fig:lab}
\end{figure}

Today, the local group operates in addition to the two accelerators a number of experimental end stations allowing experiments at the forefront of low-energy nuclear physics. The present layout of the FSU laboratory is shown in Fig.\,\ref{fig:lab}. Experiments with light radioactive ion beams, which are produced in-flight, can be performed at the \textsc{resolut} facility \citep{Wie13a}. The Array for Nuclear Astrophysics Studies with Exotic Nuclei, \textsc{anasen} \citep{Kos17a}, and the \textsc{resoneut} detector setup for resonance spectroscopy after $(d,n)$ reactions \citep{Bab18a} are major detector setups available for experiments at the \textsc{resolut} beamline. The laboratory further added to its experimental capabilities by introducing the \textsc{catrina} neutron detector array \citep{Per19a}, the MUSIC-type active target detector \textsc{encore} \citep{Ash21a}, and by installing the Super-Enge Split-Pole Spectrograph (\textsc{se-sps}) in collaboration with Louisiana State University, including its first new ancillary detector systems \textsc{sabre} \citep{Goo21a} and \textsc{cebra} \citep{Con24a} for coincidence experiments. Recently, the FSU group also installed the high-resolution $\gamma$-ray array \textsc{clarion2} and the \textsc{trinity} particle detector \citep{Gra22a} in collaboration with Oak Ridge National Laboratory. This array consists of up to 16 Compton-suppressed, Clover-type High-Purity Germanium (HPGe) detectors. 

\section{Featured experimental setups and capabilities}

\subsection{The Super-Enge Split-Pole Spectrograph (SE-SPS)}

\begin{figure}[t]
\begin{center}
\includegraphics[width=.95\linewidth]{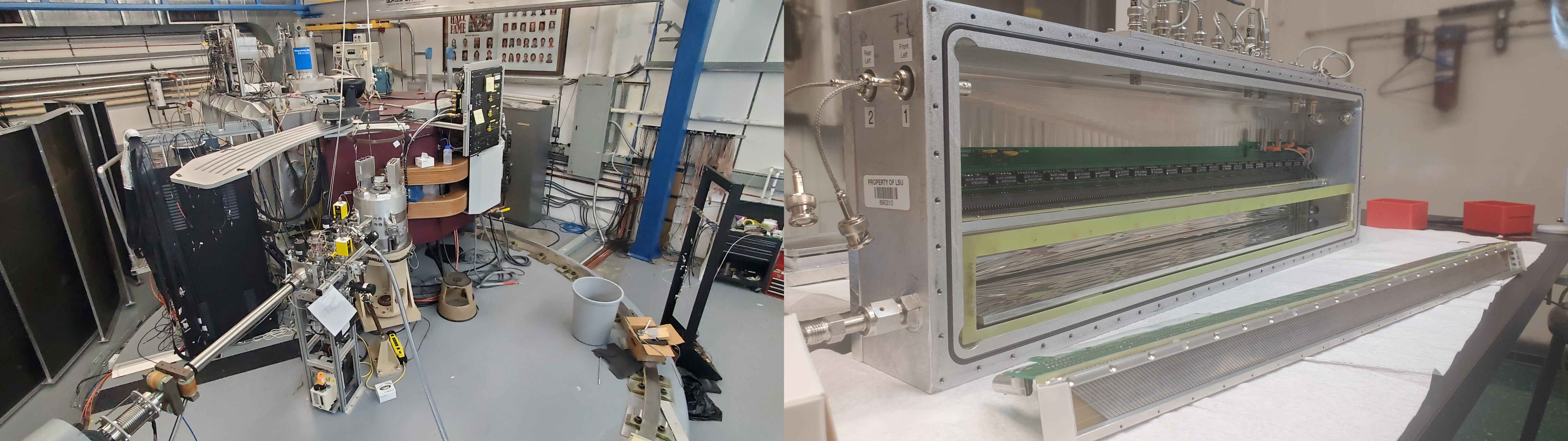}
\end{center}
\caption{The FSU Super-Enge Split-Pole Spectrograph (SE-SPS) [left]. The sliding seal scattering chamber is installed here. Parts of the rail system to measure angular distributions can be seen. The beam enters from the lower left corner. The position sensitive focal plane detector (right). The proportional-counter (tracking) section of the detector is shown and opened. The field cage and cathode plate, and some of the delay-line chips above the field-cage section can be seen on the green circuit board. One position sensitive anode wire section is taken out to show the pick-up pad structure, which is coupled to the delay-line chips and angled at $45^{\circ}$.}\label{fig:sesps}
\end{figure}

The Super-Enge Split-Pole Spectrograph (SE-SPS) has been moved to FSU after the Wright Nuclear Structure Laboratory (WNSL) at Yale University ceased operation. Like any spectrograph of the split-pole design \cite{Eng79a}, the SE-SPS consists of two pole sections used to momentum-analyze reaction products and focus them at the magnetic focal plane to identify nuclear reactions and excited states. The split-pole design allows to accomplish approximate transverse focusing as well as to maintain second-order corrections in the polar angle $\theta$ and azimuthal angle $\phi$, i.e., $(x/\theta^2) \approx 0$ and $(x/\phi^2) \approx 0$, over the entire horizontal range \cite{Eng79a}. H. Enge specifically designed the SE-SPS spectrograph as a large-acceptance modification to the traditional split-pole design for the WNSL. The increase in solid angle from 2.8 to 12.8 msr was achieved by doubling the pole-gap, making the SE-SPS well-suited for coincidence experiments. At FSU, the SE-SPS was commissioned in 2018. The design resolution of $\sim$\,20\,keV was achieved in June 2019 during a \nuc{12}{C}$(d,p)$\nuc{13}{C} experiment with a thin natural Carbon target after improvements to the accelerator optics, the dedicated beamline by adding a focusing quadrupole magnet in front of the scattering chamber, and the new CAEN digital data acquisition\,\cite{caen}. Fig.\,\ref{fig:sesps} shows the SE-SPS in target room 2 of the FSU John D. Fox Laboratory.

\begin{figure}[t]
\begin{center}
\includegraphics[width=.95\linewidth]{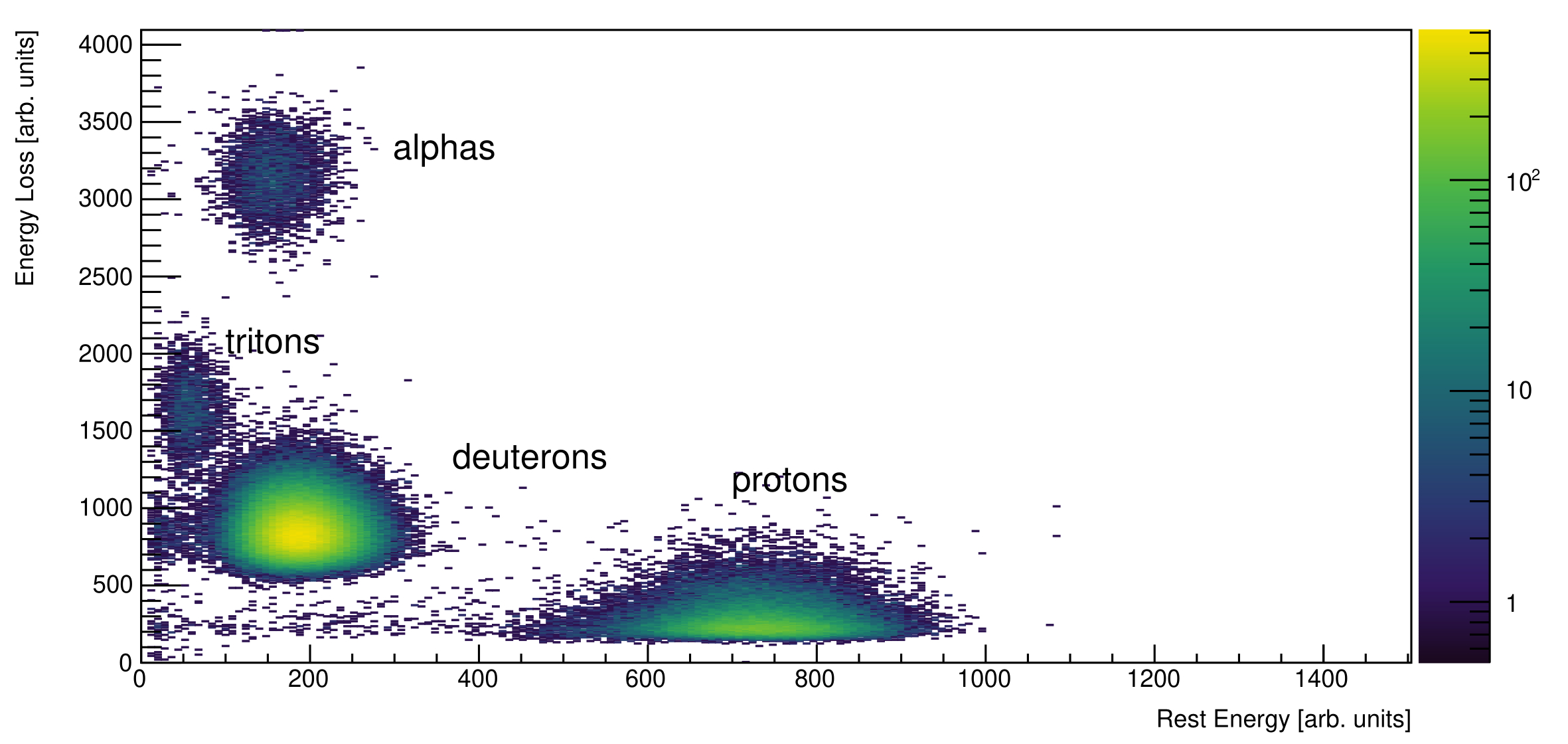}
\end{center}
\caption{Particle identification with the FSU SE-SPS. The example of deuteron-induced reactions $(d,X)$ on \nuc{49}{Ti} has been chosen. Here, protons, deuterons, tritons, and $\alpha$ particles fall within the momentum acceptance of the SE-SPS and can be clearly distinguished. The rest energy is measured with the plastic scintillator at the end of the focal-plane detector. The energy loss can be determined using one of the anode-wire signals. Here, the energy loss measured with the front-anode wire is shown.}\label{fig:sesps_pid}
\end{figure}

\begin{figure}[t]
\begin{center}
\includegraphics[width=.98\linewidth]{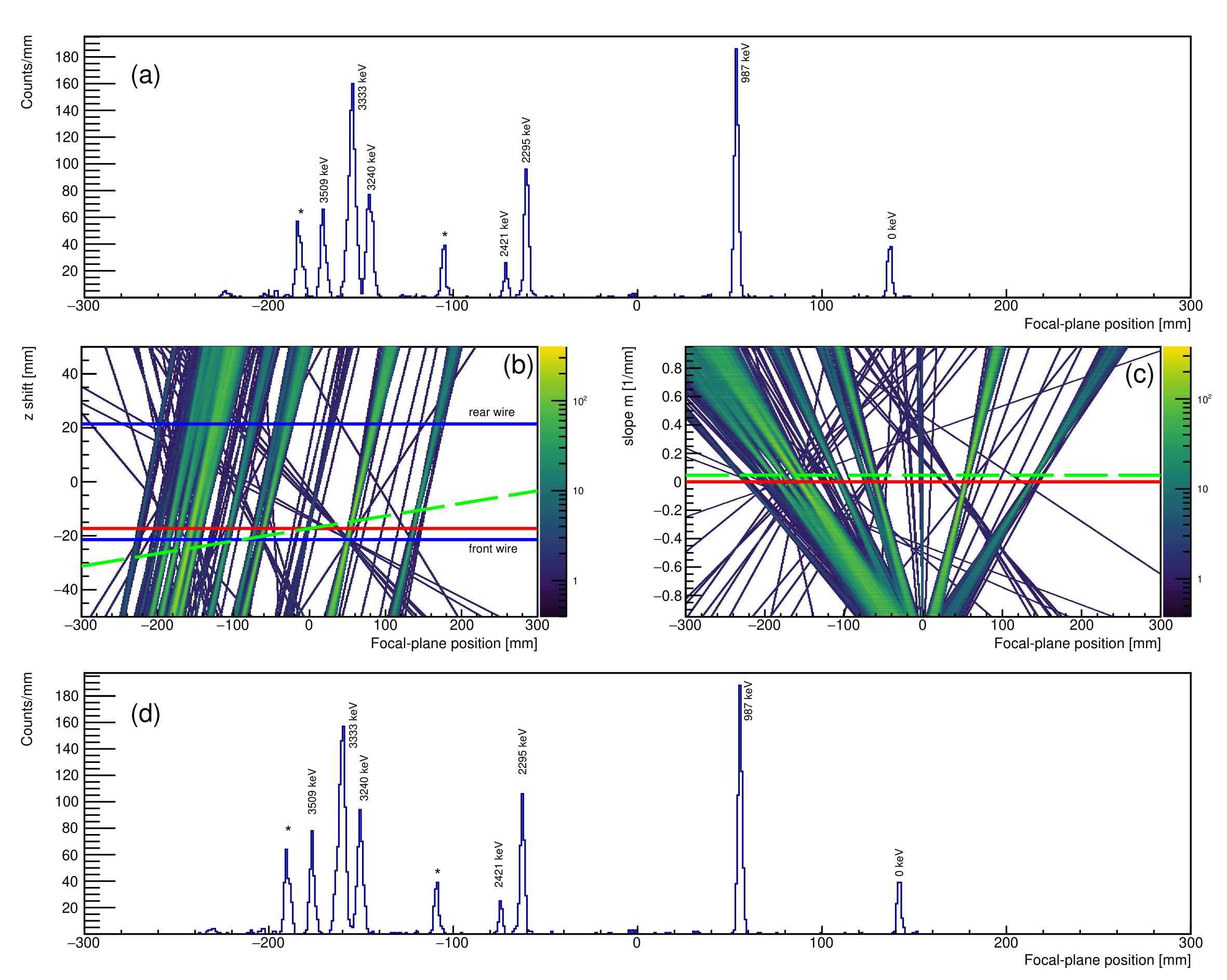}
\end{center}
\caption{(a) triton spectrum measured in $\nuc{49}{Ti}(d,t)\nuc{48}{Ti}$ with the SE-SPS placed at a laboratory scattering angle of $30^{\circ}$. Excited states of \nuc{48}{Ti} are marked with their excitation energy. Contaminants stemming from other Ti isotopes in the target are identified with asterisks. A vertical shift of the real focal plane relative to the front and back wire of the focal-plane detector was assumed [shown as red line in panels (b) and (c)]. (b) and (c) possible correction when assuming that the $z$ shift of the focal plane depends on the focal-plane position according to $z(x)=m*x+z_0$, i.e., a linear tilt. The position of the front and rear wires are highlighted with blue lines and labeled, respectively. (d) focal-plane spectrum when the linear correction of panels (b) and (c) is applied.}\label{fig:sesps_spectrum}
\end{figure}

\begin{figure}[t]
\begin{center}
\includegraphics[width=.75\linewidth]{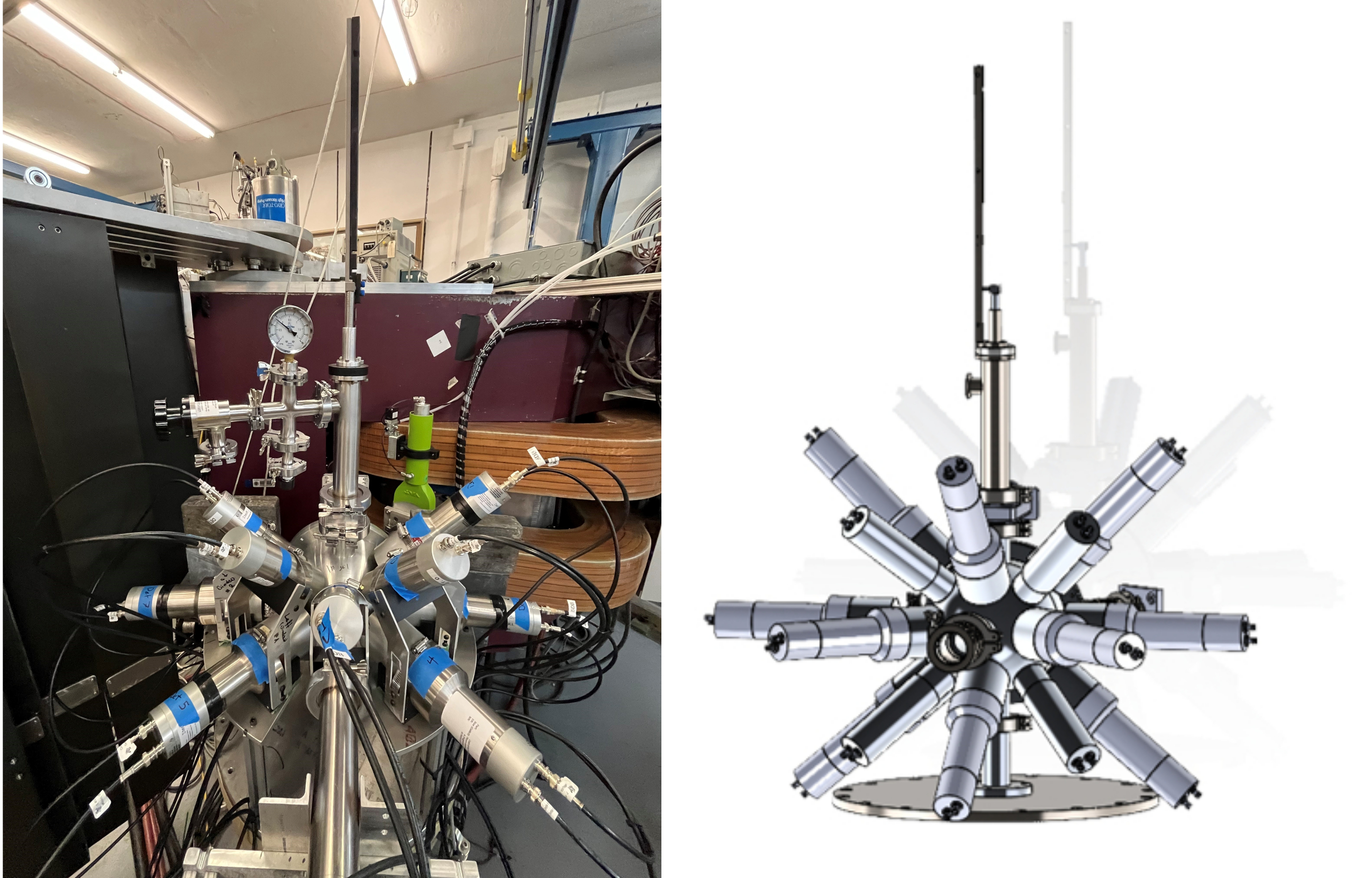}
\end{center}
\caption{The extended CeBrA demonstrator in front of the SE-SPS (left). The array consists of four $2 \times 2$ inch, four $3 \times 3$ inch, and one $3 \times 4$ inch CeBr$_3$ detectors. The $3 \times 3$ inch detectors are temporary loans from Mississippi State University. The detectors are installed around the dedicated scattering chamber. Some of the lead bricks, used to shield background coming from the Faraday cup and the SE-SPS entrance slits, can be seen. See Ref.\,\cite{Con24a} for more details on the five-detector CeBrA demonstrator. A CAD drawing of the geometry planned for the 14-detector array including four $2 \times 2$ inch, six $3 \times 4$ inch, and four $3 \times 6$ inch CeBr$_3$ detectors is also shown (right).}\label{fig:sesps_cebra}
\end{figure}

In singles experiments, i.e., stand-alone mode, the SE-SPS with its current light-ion focal plane detection system (see Fig.\,\ref{fig:sesps} and Ref.\,\cite{Goo20a}) can be used to study the population of excited states in light-ion induced reactions, determine (differential) cross sections and measure the corresponding angular distributions. Currently, laboratory scattering angles of up to $60^{\circ}$ can be covered. The focal-plane detector consists of a position-sensitive proportional counter with two anode wires (see Fig.\,\ref{fig:sesps}), separated by about 4.3\,cm, to measure position, angle, and energy loss, and a large plastic scintillator to determine the rest energy of the residual particles passing through the detector. The focal-plane detector has an active length of about 60\,cm. A sample particle identification plot with the energy loss measured by the front-anode wire and the rest energy measured by the scintillator is shown in Fig.\,\ref{fig:sesps_pid}. Unambiguous particle identification is achieved. Under favorable conditions, the detector can be operated at rates as high as two kilocounts/s (kcps). A sample position spectrum measured with the delay lines of the SE-SPS focal plane detector is also shown in Fig.\,\ref{fig:sesps_spectrum}. As the resolution depends on the solid angle, target thickness and beam-spot size, it may vary from experiment to experiment. See also comments in \cite{Eng79a, DeV77a}. In standard operation and with a global kinematic correction, i.e., assuming a vertical shift of the real focal plane with respect to the two position-sensitive sections of the detector, a full width at half maximum of 30-70\,keV has been routinely achieved. This corresponds to a position resolution of about two millimeters. This resolution can be improved further with position-dependent offline corrections. An example for such a correction, taking into account the position dependence of the $z$ shift for obtaining the true focal-plane position relative to the two anode wires and assuming that it depends linearly $z(x)=m*x+z_0$ on the focal-plane position $x$, has been added to Fig.\,\ref{fig:sesps_spectrum}. A slope of $m=0$ would correspond to the standard correction of calculating the real focal plane from a ``vertical'' shift relative to the two focal-plane wires and is shown with a red line in Figs.\,\ref{fig:sesps_spectrum}\,(b) and (c). As can be seen in Figs.\,\ref{fig:sesps_spectrum}\,(b) and (c), there is a region with $m > 0$, where the ``tracks'' mostly corresponding to excited states of \nuc{48}{Ti} populated in $\nuc{49}{Ti}(d,t)$ get narrower after the correction, thus improving the position resolution along the focal plane. The improved focal-plane spectrum is shown in Fig.\,\ref{fig:sesps_spectrum}\,(d). The magnetic field is 11.2\,kG and the solid-angle acceptance $\Delta \Omega$ was kept at 4.6\,msr for this experiment. The necessity of kinematic corrections for magnetic spectrographs and how to calculate the vertical $z$ shift for, {\it e.g.}, the split-pole design were also discussed in \cite{Eng79a}.

Angular distributions provide direct information on the angular momentum, $l$, transfer and, for one-nucleon transfer reactions, information on the involved single-particle levels. For the set of $(d,p)$ experiments performed with the SE-SPS up to date, very good to excellent agreement has been observed between the experimental data and the reaction calculations using the conventional Distorted Wave Born Approximation (DWBA) and the adiabatic distorted wave (ADW) method with input from global optical model potentials. Further details were discussed in \cite{Ril21a, Ril22a, Spi23a, Ril23a, Hay24a, Kuc24a}. Some examples for $\nuc{52}{Cr}(d,p)\nuc{53}{Cr}$ are shown in Fig.\,\ref{fig:angcorr} and will be discussed further in the next section in the context of particle-$\gamma$ coincidence experiments with the SE-SPS and CeBrA.

\subsubsection{The CeBrA demonstrator for particle-$\gamma$ coincidence experiments at the SE-SPS}

\label{sec:cebra}

\begin{figure}[t]
\begin{center}
\includegraphics[width=.85\linewidth]{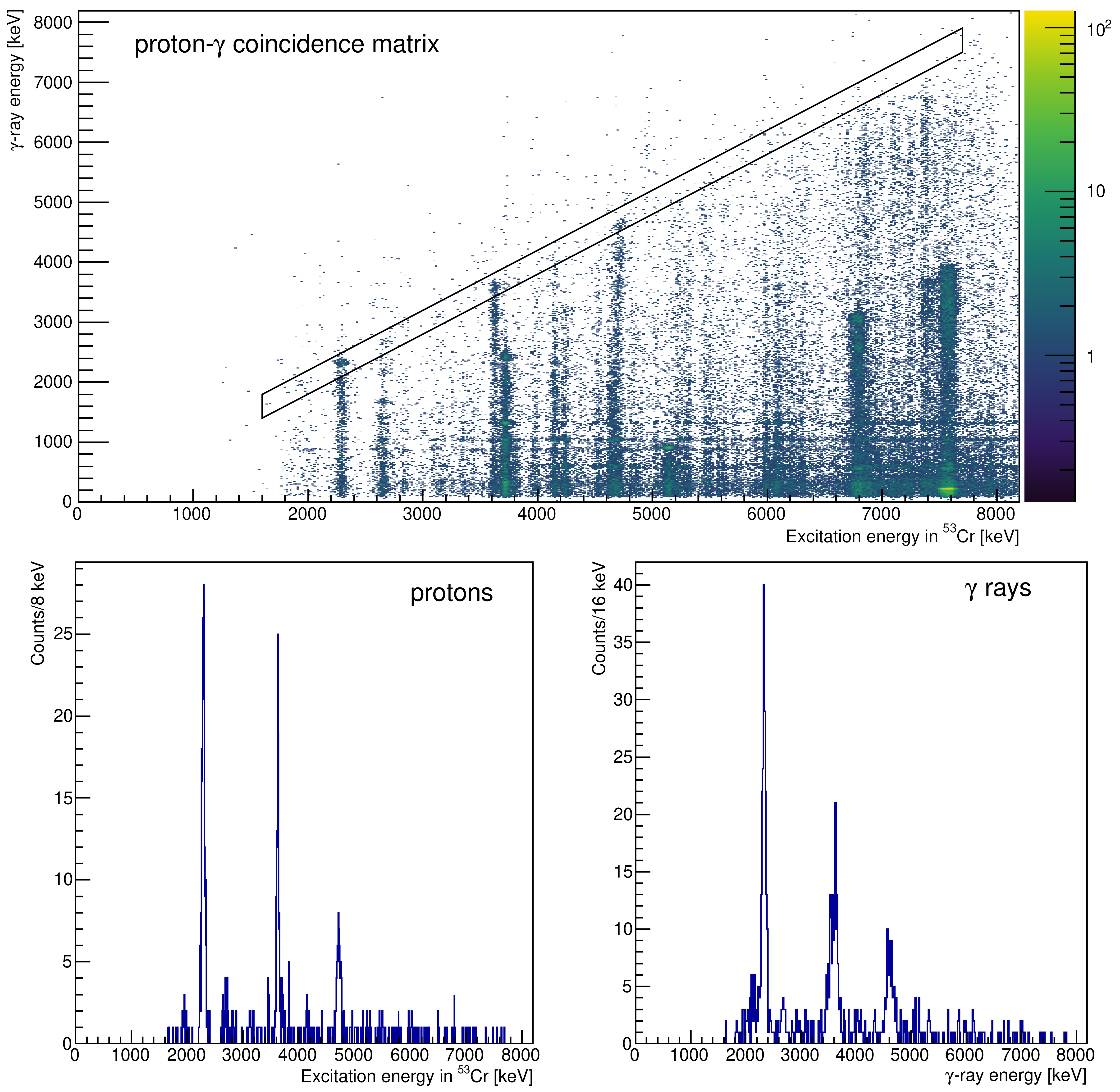}
\end{center}
\caption{Proton-$\gamma$ coincidence matrix measured in $\nuc{52}{Cr}(d,p\gamma)\nuc{53}{Cr}$ (top panel). In addition, projections onto the excitation-energy axis (protons) and onto the ``$\gamma$-ray energy'' axis ($\gamma$ rays) are shown in the two bottom panels. These spectra were obtained by applying the diagonal gate shown in the top panel to the proton-$\gamma$ coincidence matrix. This specific gate selects $\gamma$ decays to the ground state of \nuc{53}{Cr}. At higher energies, excited states of \nuc{13}{C} populated through the $\nuc{12}{C}(d,p)$ reaction on the Carbon target backing can be seen (top panel).}\label{fig:pgmatrix}
\end{figure}

The Cerium Bromide Array (CeBrA) demonstrator for particle-$\gamma$ coincidence experiments at the SE-SPS has recently been commissioned at the John D. Fox Laboratory\,\cite{Con24a}. It has been extended since with four $3 \times 3$ inch detectors on temporary loan from Mississippi State University (see Fig.\,\ref{fig:sesps_cebra}). This extended demonstrator has a combined full energy peak (FEP) efficiency of about 3.5\,$\%$ at 1.3\,MeV. For comparison, the five-detector demonstrator had an FEP efficiency of about 1.5\,$\%$ at 1.3\,MeV\,\cite{Con24a}. The comparison underscores the significant gain when adding larger volume detectors. Over the next years, a 14-detector array will be built in collaboration with Ursinus College and Ohio University through funding from the U.S. National Science Foundation, combining the existing detectors (four $2 \times 2$ inch and one $3 \times 4$ inch) of the demonstrator with five additional $3 \times 4$ inch and four $3 \times 6$ inch CeBr$_3$ detectors.

An example for a particle-$\gamma$ coincidence matrix, measured in $\nuc{52}{Cr}(d,p\gamma)\nuc{53}{Cr}$ with the five-detector demonstrator, is shown in Fig.\,\ref{fig:pgmatrix}. Using diagonal gates, $\gamma$ decays leading to specific (excited) states can be selected. In Fig.\,\ref{fig:pgmatrix}, $\gamma$ decays to the ground state of \nuc{53}{Cr} were selected. Three states stand out as they are also strongly populated in $(d,p)$\,\cite{Ril23a}. They are the excited states at 2321\,keV, 3617\,keV, and 4690\,keV. The decay of the 4690-keV state is, to our knowledge, observed for the first time. No information on its $\gamma$ decay is adopted\,\cite{ENSDF}. The $\gamma$ ray at $\sim 2.6$\,MeV indicates that, different from previous conclusions\,\cite{Ril23a}, both the 2657-keV and 2670-keV states might have been populated in $(d,p)$. The ground-state branch of the $J^{\pi} = 5/2^-$, 2657-keV state is too small to explain the excess of counts. More details will be discussed in a forthcoming publication\,\cite{Ril24a}, which will also highlight the significant value added from performing complementary singles and coincidence experiments with the SE-SPS. A feature, which can be immediately appreciated from Fig.\,\ref{fig:pgmatrix}, is that the energy resolution of the SE-SPS barely changes over the length of the focal plane, while the CeBr$_3$ energy resolution shows the expected dependence on $\gamma$-ray energy \cite{Con24a}. Using the additional $\gamma$-ray information and projecting onto the excitation-energy axis will allow us to distinguish close-lying states, which might be too close in energy to do so in SE-SPS singles experiments or where particle spectroscopy alone does not provide conclusive results. For this, differences in $\gamma$-decay behavior can be used. As an example, see the very different $\gamma$-decay behavior of the 3617-keV, $J^{\pi} =1/2^-$ and the 3707-keV, $J^{\pi}=9/2^+$ states of \nuc{53}{Cr} in Fig.\,\ref{fig:pgmatrix}. For the 3617-keV state, the 3617-keV $1/2^- \rightarrow 3/2^-_1$ ground-state transition is the strongest, while it is the 2417-keV $9/2^+ \rightarrow 7/2^-_1$ transition for the 3707-keV state. Another example, using different diagonal gates for the $\nuc{61}{Ni}(d,p\gamma)\nuc{62}{Ni}$ reaction and, thus, selecting $\gamma$ decays leading to different final states with different $J^{\pi}$ as ``spin filter'', was featured in Ref.\,\cite{Con24a}.

\begin{figure}[t]
\begin{center}
\includegraphics[width=.95\linewidth]{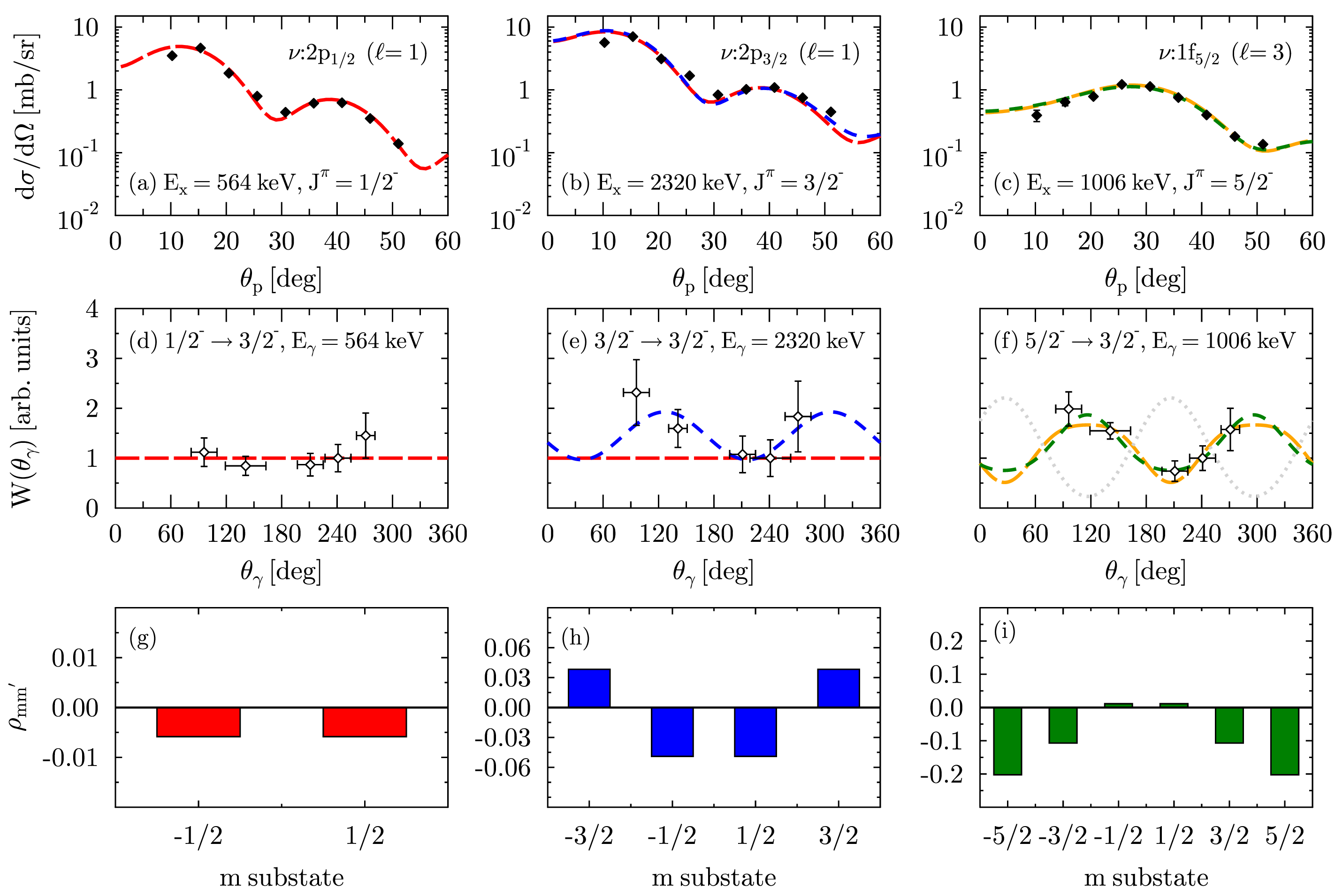}
\end{center}
\caption{(a)--(c) $\nuc{52}{Cr}(d,p)\nuc{53}{Cr}$ angular distributions and (d)--(f) proton-$\gamma$ angular correlations measured in $\nuc{52}{Cr}(d,p\gamma)\nuc{53}{Cr}$ for the 564-keV, $J^{\pi}=1/2^-$ state, the 2320-keV, $J^{\pi}=3/2^-$ state, and the 1006-keV, $J^{\pi}=5/2^-$ state. The angular correlations for (d) the 564-keV, $1/2^- \rightarrow 3/2^-_1$, (e) 2320-keV, $3/2^- \rightarrow 3/2^-_1$, and (f) 1006-keV, $5/2^- \rightarrow 3/2^-_1$ $\gamma$-ray ground-state transitions are shown, respectively. In addition, predictions from (a)--(c) ADW calculations with \textsc{chuck3}\,\cite{CHUCK}, and (e)--(f) combined ADW calculations and \textsc{angcor}\,\cite{angcor} calculations to generate the angular correlations are shown for each transition (lines). (g)--(i) Density matrices $\rho_{mm^{\prime}}$ as defined in, {\it e.g.}, Ref.\,\cite{Ryb70a}. The proton-$\gamma$ angular correlation for the $5/2^- \rightarrow 3/2^-$ transition calculated with the currently adopted multipole-mixing ratio of $\delta = 0.36(2)$ was added to (f) [gray, dotted line]. Different sign conventions for the multipole mixing ratio are likely the origin of the disagreement. Note that the $y$-scale in panels (d)--(f) is the same.}\label{fig:angcorr}
\end{figure}

The coincidently detected $\gamma$ rays also provide access to important complementary information such as $\gamma$-decay branching ratios and particle-$\gamma$ angular correlations for spin-parity assignments, as well as the possibility to determine nuclear level lifetimes via fast-timing techniques and excluding feeding due to gates on the excitation energy\,\cite{Con24a}. For the latter, the smaller detectors are better suited because of their better intrinsic timing resolution as also discussed in Ref.\,\cite{Con24a}. For dedicated fast-timing measurements, two $1 \times 1$ inch CeBr$_3$ detectors are available at FSU in addition to the four $2 \times 2$ inch detectors. These have an even better timing resolution than the $2 \times 2$ inch detectors, however, at the cost of a significantly lower FEP efficiency. A careful analysis of their timing properties and FEP efficiencies is ongoing.

We will briefly highlight some particle-$\gamma$ angular correlations measured with the five detector CeBrA demonstrator. Particularly, we will discuss how these can be used to make spin-parity assignments and to determine multipole mixing ratios $\delta$. Fig.\,\ref{fig:angcorr} shows three proton-$\gamma$ angular correlations measured in $\nuc{52}{Cr}(d,p\gamma)\nuc{53}{Cr}$ and with all five CeBr$_3$ detectors placed in a common plane with an azimuthal angle $\phi_{\gamma} = 0^{\circ}$. In addition to the experimental data, predictions from combined ADW calculations with \textsc{chuck3}\,\cite{CHUCK} yielding scattering amplitudes and \textsc{angcor}\,\cite{angcor} calculations using these scattering amplitudes to generate the angular correlations are shown. The associated density matrices, $\rho_{mm^{\prime}}$, needed to calculate the proton-$\gamma$ angular correlations with the formalism presented in Ref.\,\cite{Ryb70a} and which are connected to the scattering amplitudes for the different $m$ substates, were added, too. As all the $\gamma$-ray transitions of Fig.\,\ref{fig:angcorr} are primary transitions, the multipole mixing ratio $\delta$ is the only free parameter. It was determined via $\chi^2$ minimization. Excellent agreement is observed between the experimentally measured and the calculated distributions for the excited states at $E_x = 564$\,keV, 1006\,keV, and 2320\,keV of \nuc{53}{Cr}. For the 564-keV state, a one-neutron transfer to the $2p_{1/2}$ neutron orbital was assumed [red, longer dashed line in Fig.\,\ref{fig:angcorr}\,(a)]. For the 2320-keV state, the neutron was transferred into the $2p_{3/2}$ orbital (blue, shorter dashed line). For the 1006-keV state, the neutron was transferred into the the $1f_{5/2}$ orbital (green, shorter dashed line). For the 2320-keV and 1006-keV states, transfers to their corresponding spin-orbit partner are also shown in Fig.\,\ref{fig:angcorr}. In panels (c) and (f), predictions for a neutron transfer into the $1f_{7/2}$ orbital are shown with orange, longer dashed lines. 

As expected for the 564-keV, $1/2^-_1 \rightarrow 3/2^-_1$ ground-state transition, the negligible alignment [see Fig.\,\ref{fig:angcorr}\,(g)] leads to an isotropic angular distribution. We note that this is true for any value of $\delta$. For the 2320-keV, $3/2^- \rightarrow 3/2^-_1$ and 1006-keV, $5/2^-_1 \rightarrow 3/2^-_1$ ground-state transitions, the $m$-substate population (alignment) [see Figs.\,\ref{fig:angcorr}\,(h) and (i)] results in observable angular distributions. In both cases, the multipole mixing ratio indicates that the transition is dominantly of $E2$ character. A more in depth discussion will be provided in a forthcoming publication\,\cite{Ril24a}. Figs.\,\ref{fig:angcorr}\,(d) and (e) show clearly that $J^{\pi} = 1/2^-$ and $J^{\pi} = 3/2^-$ states can be distinguished based on their observed proton-$\gamma$ angular correlation. $(d,p)$ singles experiments with an unpolarized deuteron beam cannot discriminate between these states since both are populated via an $l=1$ angular momentum transfer [see Figs.\,\ref{fig:angcorr}\,(a) and (b)]. The situation appears more complex for the $f$ orbitals, where the predicted proton-$\gamma$ angular correlations are not sufficiently different to discriminate between a $7/2^- \rightarrow 3/2^-$ and $5/2^- \rightarrow 3/2^-$ transition [see Fig.\,\ref{fig:angcorr}\,(f)]. For the known 1006-keV, $J^{\pi} = 5/2^-$ state, the calculation assuming a neutron transfer into the $1f_{5/2}$ orbital does provide the slightly better $\chi^2$ value though. For completeness, we added the proton-$\gamma$ angular correlation for the $5/2^- \rightarrow 3/2^-$ transition calculated with the currently adopted multipole-mixing ratio to Fig.\,\ref{fig:angcorr}\,(f). As discussed in \cite{Con24a}, the adopted ratio appears to be incorrect. However, different sign conventions for the multipole mixing ratio could also be the origin of the disagreement.

With more detectors, which will be added to the full CeBrA array within the next couple of years, statistics will increase and particle-$\gamma$ angular correlation measurements can be performed in planes with varying $\phi_{\gamma}$. Four ``rings'' will be available in the standard configuration, where three of them have at least four detectors (see Fig.\,\ref{fig:sesps_cebra}). The full setup will allow to further test details of different transfer reactions and the predicted alignment. Measuring particle-$\gamma$ angular correlations in planes with different $\theta_{\gamma}$ could potentially help to better discriminate between spin-orbit partners, like $1f_{5/2}$ and $1f_{7/2}$, too. Another example of how different angular correlations can look for the $2d_{5/2}$ and $2d_{3/2}$ spin-orbit partners will be shown in Sec.\,\ref{sec:pdr}.

\subsection{The CATRiNA neutron detector array}

\begin{figure}
\begin{center}
\includegraphics[width=.95\linewidth]{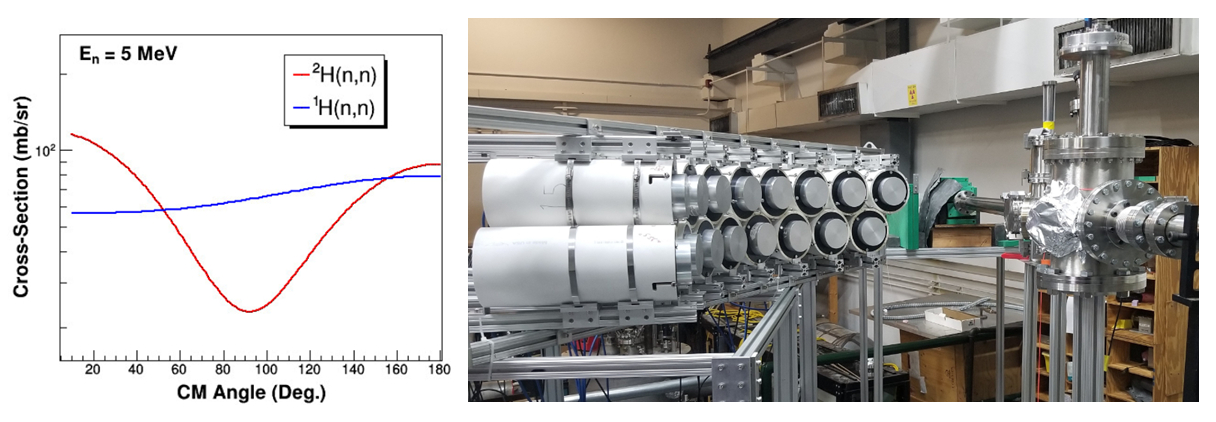}
\end{center}
\caption{(left) DWBA calculations made with the \textsc{fresco} computer program \cite{Tho88a} for $n-d$, and $n-p$ elastic scattering showing the difference between the isotropic angular distribution for $n-p$ scattering and the non-isotropic angular distribution for $n-d$ scattering. (right) The CATRiNA neutron detector array in target room \#1 at the John D. Fox Laboratory. }\label{fig:Ang-setup}
\end{figure}

The CATRiNA neutron detector array currently consists of 32 deuterated-benzene (C$_6$D$_6$) liquid scintillator neutron detectors. There are two sizes of CATRiNA detectors: 16 ``small'' detectors and 16 ``large'' detectors. The ``small'' CATRiNA detectors encapsulate the deuterated scintillating material in a 2'' diameter $\times$ 2'' deep cylindrical aluminum cell, while the ``large'' detectors encapsulate the scintillating material in a 4'' diameter $\times$ 2'' deep cylindrical aluminum cell \cite{perello-nim, morelock-nim}.

The use of deuterated scintillating material for neutron detection, rather than traditional hydrogen-based scintillating material, is due to unique features produced in the light-output spectrum. Neutrons scattered off the deuterium in the scintillator will produce a characteristic forward recoil peak and low valley in the light-output spectrum. This feature is due to the asymmetry of the cross section for $n-d$ scattering, which peaks at backwards angles and extends across a large range of neutron energies. As an example, Fig. \ref{fig:Ang-setup} shows a DWBA calculation made with the \textsc{fresco} computer program \cite{Tho88a} for the elastic scattering cross sections of 5-MeV neutrons off the deuteron \nuc{2}{H} and proton \nuc{1}{H} as a function of the center-of-mass (CM) angle. The difference between the angular distributions can be clearly seen. The characteristic light-output spectra of deuterated scintillators is then used for the extraction of neutron energies using spectrum-unfolding methods. Determining neutron energies from spectrum unfolding is an alternative to fully relying on time-of-flight (ToF) information for neutron energies. This alternative is  particularly beneficial if a compact neutron detector system like CATRiNA is used, which efficiently optimizes solid angle coverage and the size of the detector array for neutron studies. The CATRiNA detectors have equivalent properties of organic scintillators for neutron detection such as large scattering cross section for neutrons with the scintillating material, fast response time, and pulse shape discrimination capabilities that allow separation of neutron ($n$) and gamma-ray ($\gamma$) events. 

To highlight the capabilities of the CATRiNA neutron detectors, a $(d,n)$ proton-transfer experiment was conducted on a solid deuterated-polyethylene, CD$_2$, target of 400-$\mu$g/cm$^2$ thickness and a set of ``large'' CATRiNA detectors placed in target room \#1 of the Fox Laboratory. The FN Tandem accelerator provided deuteron beams with energy $E_d = 5 - 8$\,MeV. The deuteron beam was bunched to 2-ns width with intervals of 82.5 ns for time-of-flight (ToF) measurements using the accelerator’s radiofrequency (RF) as reference signal. The CATRiNA detectors were placed at 1-m distance from the CD$_2$ target. A thick graphite disk was placed 2 m downstream from the target and used as a beam stop to minimize beam-induced background. The graphite beam stop was held inside a 30 cm $\times$ 30 cm $\times$ 30 cm borated-polyethylene block, which was surrounded with 5-cm thick lead bricks and thin lead sheets to reduce background from beam-induced neutrons and $\gamma$ rays from the beamstop, respectively. The experimental setup is shown in Fig. \ref{fig:Ang-setup}. The characterization of the CATRiNA detectors and the description of the unfolding method can be found in Refs.\,\cite{perello-nim, morelock-nim}. 
Neutrons from the interaction of the deuterium beam with the carbon and deuterium in the CD$_2$ target were used to compare neutron energies measured with ToF and extracted with the unfolding methods. 

\begin{figure}
\begin{center}
\includegraphics[width=.95\linewidth]{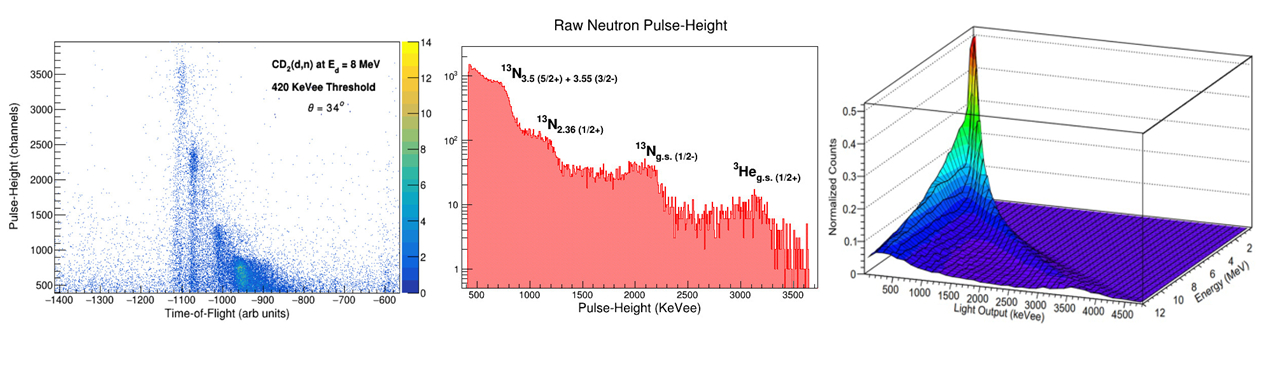}
\end{center}
\caption{(left) Pulse-height vs. ToF correlation for neutrons from the interaction of an 8-MeV deuteron beam with a 400-$\mu$m/cm$^2$ thick CD$_2$ target. (middle) Raw pulse-height spectrum obtained from projecting neutron events in a PSD plot on the long integration axis. Different neutron groups can be identified. (right) Simulated response matrix for the CATRiNA detectors. The simulation was performed using the Monte Carlo neutron transfer code MCNP6\,\cite{mcnp6}.}\label{fig:unfolding}
\end{figure}

In the following, the pulse-shape discrimination (PSD) properties of the CATRiNA detectors have been used to separate neutron and $\gamma$-ray interactions in the detectors. For the ToF measurements, the time difference was measured between the prompt-gamma signal and the neutron peaks coming from the interaction of the beam with the target. The accelerator’s RF signal was used as a ``stop'' signal while the ``start'' signal was provided by an``or'' of any events registered in the CATRiNA detectors. The energy of the neutrons was then calculated using non-relativistic kinematics taking into account the target to detector distance and the measured time of flight.

\begin{figure}
\begin{center}
\includegraphics[width=.95\linewidth]{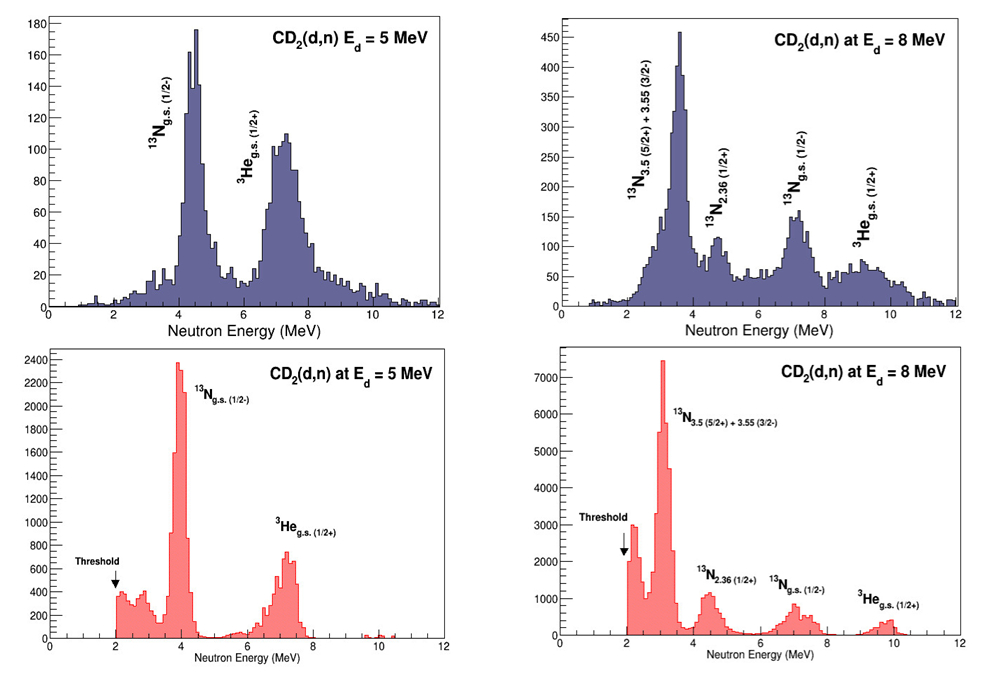}
\end{center}
\caption{Direct comparison of the neutron-energy spectra obtained via time-of-flight (top panels) versus those obtained with an unfolding method (bottom panels). Data obtained from $(d,n)$ reactions with deuteron-beam energies of $E_d = 5$\,MeV and 8 MeV are shown.}\label{fig:comparison}
\end{figure}

For the extraction of neutron energies via unfolding, the pulse height spectrum was analyzed. The raw pulse-height spectra of the detectors is obtained by gating the neutron events in the PSD plots and projecting onto the long-integration axis. A correlation matrix, ToF vs. pulse height, of neutron events from interaction of an 8-MeV deuteron beam with the CD$_2$ target is shown in Fig.\,\ref{fig:unfolding}. It can be seen that the most energetic neutrons have the highest pulse-height values. The raw pulse-height spectra show distinctive shoulders that shift to the right as the neutron energy increases and can be attributed to separate states populated in the reaction. A typical raw pulse height spectrum is shown in Fig.\,\ref{fig:unfolding}. 

To unfold the neutron energies, a response matrix needs to be created. The response matrix correlates the light-output (or pulse-height) spectra of the detectors with the neutron energies and the detector efficiencies. A statistical method is then employed to extract energies of incident neutrons by comparing to the response matrix of the detector in an iterative process. The present data was analyzed using a response matrix simulated with the Monte Carlo neutron-particle transport code MCNP6 \cite{mcnp6} and validated using selected mono-energetic neutrons from the $\nuc{7}{Li}(p,n)$ reaction \cite{perello-nim, perello-thesis}. The response matrix for one of the ``large'' CATRiNA detectors is shown in Fig.\,\ref{fig:unfolding}. The neutron energies extracted via unfolding method were obtained using a statistical algorithm with the Maximum-Likelihood Expectation Method (MLEM) \cite{perello-nim, perello-thesis}. Neutron energies obtained by the described spectrum unfolding method were compared with the neutron energies obtained from the ToF method. States in $^{13}$N were populated by the $^{12}$C(d,n)$^{13}$N reaction. At $E_d = 5$\,MeV, the energy of neutrons corresponding to the population of the $1/2^-$ ground state in $^{13}$N is around 4\,MeV for the angles measured. Similarly, the $1/2^+$ ground state in $^{3}$He populated by the $^2$H$(d,n){}^3$He reaction is visible at $\sim 7$\,MeV. A software threshold cut of around 2\,MeV was placed on the neutron energies to minimize neutron background and obtain a clean $n/\gamma$ separation with the CATRiNA detectors. The neutron spectra for a detector placed at 34$^{\circ}$ obtained by both methods is shown in Fig.\,\ref{fig:comparison}. As the beam energy was increased, other features of the spectrum became visible. At $E_d = 8$\,MeV, neutrons corresponding to the $J^{\pi} = 1/2^-$ g.s in $^{13}$N have neutron energies of around 7.2\,MeV. In addition, neutrons corresponding to the population of the $1/2^+$ first excited state in $^{13}$N at $E_{ex}$ = 2.36 MeV are detected at 4.8\,MeV, and a doublet with spin-parity assignments of $3/2^{-}$ and $5/2^{+}$, respectively, and $E_{x} \approx 3.5$\,MeV is observed at 3.6\,MeV. The $1/2^+$ ground state in $^{3}$He is now visible at around 9.5\,MeV.

\begin{figure}
\begin{center}
\includegraphics[width=.95\linewidth]{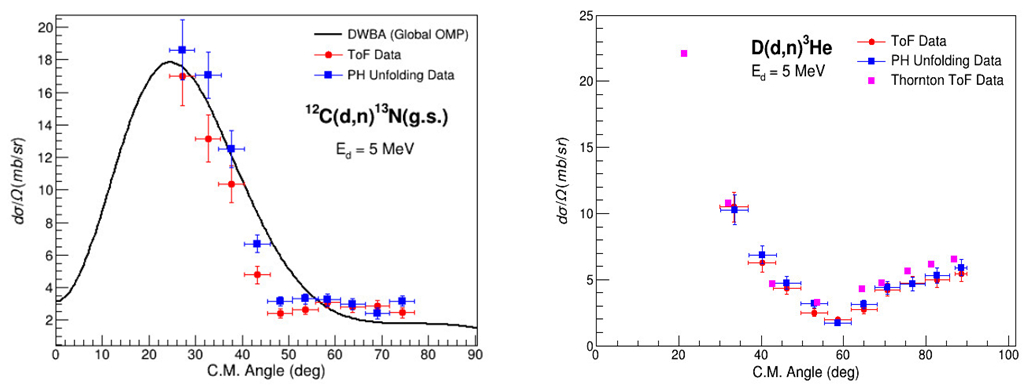}
\end{center}
\caption{Angular distributions obtained from the interaction of a 5-MeV deuteron beam with the CD$_2$ target. DWBA calculations were made with \textsc{fresco}\,\cite{Tho88a}. Angular distributions obtained with the ToF and unfolding method are compared showing excellent agreement.}\label{fig:spectroscopic}
\end{figure}

The direct comparison of neutron spectra obtained by ToF and by unfolding procedures in Fig.\,\ref{fig:comparison} shows the potential of the CATRiNA detectors. Since the commissioning experiment reported here, the unfolding method has been improved with better experimental response matrices, which initially limited the resolution of the CATRiNA detectors. A Novel Unfolding algorithm Using Bayesian Iterative Statistics (\textsc{anubis}) was developed. ANUBIS takes into account uncertainties associated with the unfolding algorithm and determines stopping criteria to optimize the procedure \cite{morelock-nim}. Angular distributions from the $\nuc{12}{C}(d,n)\nuc{13}{N}_{gs}$ and from the $^{2}$H$(d,n){}^{3}$He reactions using a 5-MeV deuteron beam are shown in Fig.\,\ref{fig:spectroscopic}. Comparison between the angular distributions with ToF and unfolding methods are in very good agreement, additionally validating the two independent approaches.

CATRiNA is envisioned to play a central role at the John D. Fox Laboratory for neutron spectroscopy studies as well as for coincidence measurements between neutrons, $\gamma$ rays, and charged particles using the different detector systems available at the laboratory.

\subsection{The CLARION2+TRINITY array for high-resolution $\gamma$-ray spectroscopy and reaction-channel selection}

CLARION2-TRINITY is a new setup at the John D. Fox Laboratory for high-resolution $\gamma$-ray spectroscopy in conjunction with charged particle detection\,\cite{Gra22a}. The $\gamma$ rays are recorded by Clover-type High-Purity Germanium detectors (HPGe) detectors. The geometry is chosen to be non-Archimedian and detectors are arranged such that no detectors have a separation of $\Delta \theta = 180^{\circ}$ to suppress coincident detection of 511-keV $\gamma$ rays from pair production. The TRINITY particle detector uses a relatively new type of scintillator, Gadolinium Aluminum Gallium Garnet (Gd$_3$Al$_2$Ga$_3$O$_{12}$) doped with Cerium (GAGG:Ce). This scintillator has intrinsic particle discrimination capabilities through two decay components with different decay times and varying relative amplitudes. The particle identification with the GAGG:Ce is obtained by comparing waveform integrals of the fast ``peak'' and the delayed ``tail''. The ratio of these two quantities allows to discriminate between protons, $\alpha$ particles, and heavier ions. The array was commissioned in December 2021 with nine clover-type HPGe detectors and two rings of GAGG:Ce scintillators\, \cite{Gra22a}. This initial setup has now been augmented with a tenth clover-type HPGe detector and all five GAGG:Ce rings of TRINITY installed. More details on the combined setup including a description of energy-loss and contaminant measurements with the zero-degree GAGG:Ce detector can be found in \cite{Gra22a}. The first science publication from the array features results from the safe Coulomb excitation of Ti isotopes and focuses on the suppression of quadrupole collectivity in \nuc{49}{Ti}\,\cite{Gra24a}.

\begin{figure}[t]
\begin{center}
\includegraphics[width=.95\linewidth]{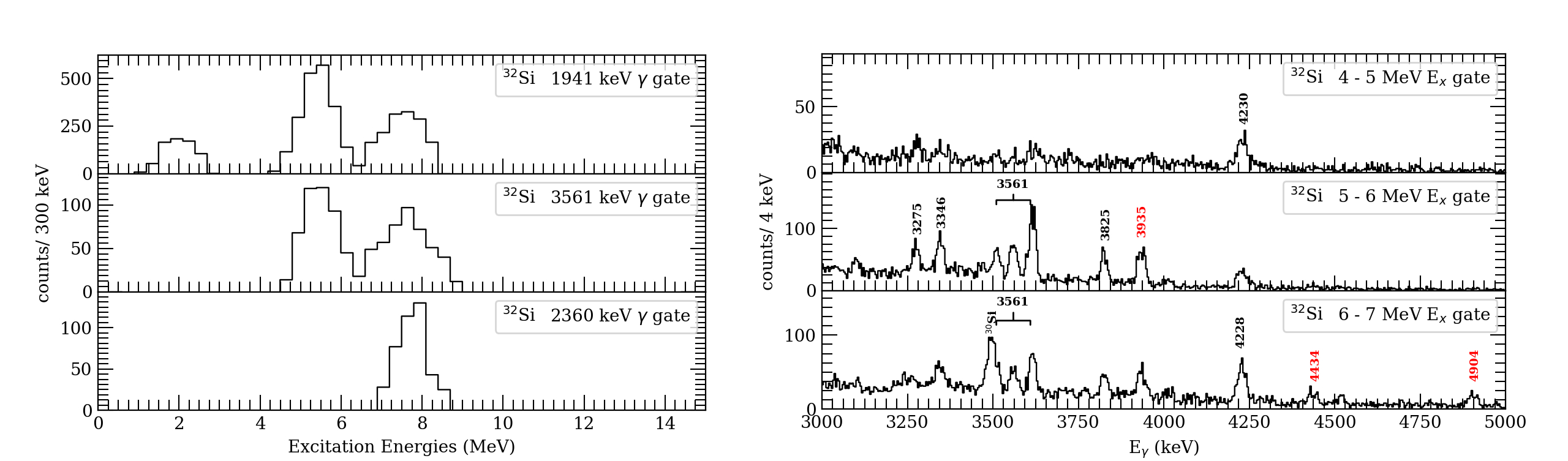}
\end{center}
\caption{$\nuc{16}{O}(\nuc{18}{O},2p)\nuc{32}{Si}$ reaction measured with CLARION2+TRINITY and a beam energy of 30\,MeV. (left) Reconstructed excitation energy spectra of \nuc{32}{Si} when gating on specific $\gamma$-ray transitions deteced with the CLARION2 Clover detectors. (right) Excitation energy gated $\gamma$-ray spectra for three 1-MeV wide, excitation-energy windows. $\gamma$-ray transitions marked in red were observed for the first time. See text for more information.}\label{fig:32si_clarion2}
\end{figure}

The setup has also been used to study unstable \nuc{32}{Si} in the $\nuc{16}{O}(\nuc{18}{O},2p)\nuc{32}{Si}$ fusion-evaporation reaction. The weak $2p$ evaporation channel could be isolated selectively by detecting both protons with TRINITY. For this reaction, triple coincidences between the two protons and $\gamma$ rays were detected with CLARION2+TRINITY. As the beam energy is precisely known and the setup allows to measure the energies and angles of the outgoing protons, the excitation energy in \nuc{32}{Si}, from which $\gamma$ rays were emitted, as well as the velocity and direction of the \nuc{32}{Si} recoil at the time of emission of the $\gamma$ ray could be reconstructed. As the ``complete'' kinematics of the reaction are known, excitation-energy gated $\gamma$-ray spectra as well as $\gamma$-transition gated excitation-energy spectra could be generated (see Fig.\,\ref{fig:32si_clarion2} for an example). As can be seen in Fig.\,\ref{fig:32si_clarion2}, the combined CLARION2+TRINITY system provides high resolution for $\gamma$ rays and moderate resolution in the excitation-energy spectra, mainly due to the target thickness and limited energy resolution of the GAGG:Ce scintillators of TRINITY. For the $\nuc{16}{O}(\nuc{18}{O},2p)\nuc{32}{Si}$ reaction, which is a weak reaction channel, excitation-energy gating provided considerably better statistics for angular distribution and polarization analysis of $\gamma$-ray transitions than a conventional $\gamma \gamma$-coincidence analysis. Some details of the reaction-channel selection were already discussed in \cite{Gra22a}. More details and results will be presented in a forthcoming publication.

\section{Selected science highlights (2020-2024)}

\subsection{Single-particle strengths around $N=28$ measured with the SE-SPS}

Spectroscopic factors obtained from one-nucleon adding and removal reactions have been critically discussed in recent years, especially for rare isotopes with large proton to neutron separation energy asymmetries (see, {\it e.g.}, Refs.\,\cite{Lee10a,Tos21a,Kay22a} and references therein). In stable nuclei, it is commonly accepted that only about 60\,$\%$ of the predicted spectroscopic strengths are observed experimentally (see, {\it e.g.}, compilations in Refs.\,\cite{Tos21a,Kay22a,Lee07a,Lee09a,Tsa09a}). Often, systematics are, however, only available for a few selected nuclei, a few isotopic or isotonic chains, and for the spectroscopic strength of a specific single-particle orbit. 

\begin{figure}[t]
\begin{center}
\includegraphics[width=.6\linewidth]{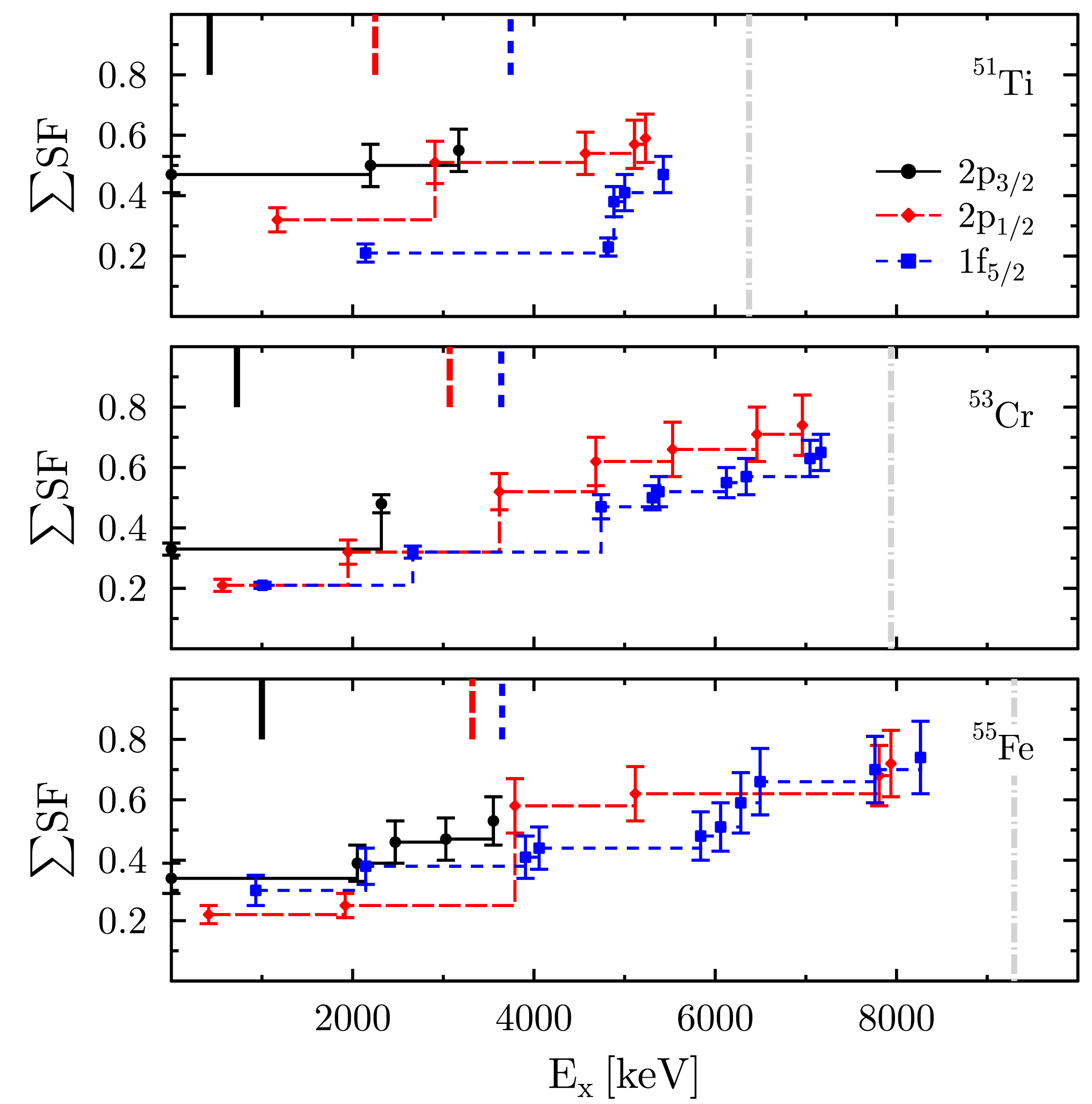}
\end{center}
\caption{Running sum of the spectroscopic strengths for the neutron $2p_{3/2}$ (black circles), $2p_{1/2}$ (red triangles), and $1f_{5/2}$ (blue squares) orbitals measured for the even-$Z$, $N=29$ isotones \nuc{51}{Ti}\,\cite{Ril21a}, \nuc{55}{Fe}\,\cite{Ril22a}, and \nuc{53}{Cr}\,\cite{Ril23a}. The centroid energies reported in Ref.\,\cite{Ril21a, Ril22a, Ril23a} are also shown with vertical bars of the corresponding colors. Uncertainties were discussed in \cite{Ril21a, Ril22a, Ril23a}. The gray dashed line corresponds to the neutron-separation energy of the corresponding nucleus. Measurements were performed up to that energy.}\label{fig:n29}
\end{figure}

In Fig.\,\ref{fig:n29}, we show a systematic study of the running sum for the neutron spectroscopic factors $\mathrm{SF} = \sigma_{exp.}/\sigma_{s.p.}$ for the even-$Z$, $N=29$ isotones; $\sigma_{s.p.}$ is the single particle cross section predicted for an excited state with excitation energy $E_x$ from ADW calculations. The $N=29$ isotones were studied at the FSU SE-SPS in $(d,p)$ experiments\,\cite{Ril21a, Ril22a, Ril23a}. As can be seen, about $50-70$\,$\%$ of the expected strength are exhausted in all three nuclei and for all three single-particle orbitals. However, it is also quite clear that it is not sufficient to just study the first few excited states. Significant parts of the $2p_{3/2}$, $2p_{1/2}$ and $1f_{5/2}$ spectroscopic strengths are fragmented to excited states with higher excitation energies. Especially for the $2p_{1/2}$ and $1f_{5/2}$ strengths, the strength is fragmented among excited states up to the neutron-separation energy, $S_n$. Studying the fragmentation of the spectroscopic strengths in $(d,p)$ experiments up to such high energies allows for a more reliable extraction of the centroid energies of the neutron single-particle orbitals. It should be noted though, that, if orbitals were partially filled, one would in general need to perform both the adding and removal reactions to experimentally determine occupancies and the real single-particle orbital energies (see, {\it e.g.}, \cite{Sch12a, Sch13a} and comments therein).

With our new data on the energies of the single-particle orbitals, we could address the disappearance of the $N=32$ and $N=34$ subshell gaps in the heavier isotones. The $N=32$ subshell gap for Ca and Ti isotopes, and its disappearance in Cr and Fe isotopes were discussed previously (see, {\it e.g.}, \cite{Pri01a,Lei18a} and references therein). In Ref.\,\cite{Ril21a}, it was stated that the closure of the $N=32$ subshell gap in the transition from Ti to Cr would need to be explained by the placement of the $1f_{5/2}$ neutron orbit relative to the $2p_{1/2}$ orbit. Within the remaining uncertainties discussed in \cite{Ril21a, Ril22a, Ril23a}, our recent $(d,p)$ studies indeed support that the gap between these two orbits shrinks with increasing proton number (see Fig.\,\ref{fig:n29}), possibly explaining the closing of the $N=32$ subshell gap in heavier isotones. The data do, however, also show that rather than the $1f_{5/2}$ centroid coming significantly down in energy, it is the $2p_{1/2}$ orbital's centroid energy which increases. This is different from the initial hypothesis\,\cite{Ril21a} and underlines the importance of performing systematic studies of spectroscopic strengths along isotopic and isotonic chains. The disappearance of the gap between the $1f_{5/2}$ and $2p_{1/2}$ neutron orbits with increasing proton number might also explain the possibly very localized occurrence of the $N=34$ subshell gap (see \cite{Liu19a} and references therein).

\subsection{The neutron one-particle-one-hole structure of the pygmy dipole resonance}

\label{sec:pdr}

The pygmy dipole resonance (PDR) has been observed on the low-energy tail of the isovector giant dipole resonance (IVGDR) below and above the neutron-separation threshold. While the additional strength is recognized as a feature of the electric dipole response of many nuclei with neutron excess\,\cite{Sav13a, Bra19a, Lan23a}, its microscopic structure, which intimately determines its contribution to the overall strength, is still poorly understood making reliable predictions of the PDR in neutron-rich nuclei far off stability difficult. It has been shown that the coupling to complex configurations drives the strength fragmentation for both the IVGDR and the PDR, and that more strength gets fragmented to lower energies when including such configurations (see the review article\,\cite{Lan23a}). The wavefunctions of $J^{\pi}=1^-$ states belonging to the PDR are, however, expected to be dominated by one-particle-one-hole (1p-1h) excitations of the excess neutrons. First experiments were performed to access these parts of the wavefunction via inelastic proton scattering through isobaric analog resonances and via one-neutron transfer $(d,p)$ experiments. The experimental results were compared to predictions from large scale shell model calculations including up to two-particle-two-hole (2p-2h) excitations for both protons and neutrons, and to quasiparticle phonon model (QPM) calculations including up to 3-phonon excitations. The comparison of experiment and theory for doubly-magic \nuc{208}{Pb} \cite{Spi20a} and semi-magic \nuc{120}{Sn} \cite{Wei21a} indicates that PDR states' wavefunctions are indeed largely dominated by 1p-1h excitations of the excess neutrons. It is important to note that $(d,p)$ experiments are not able to access all relevant neutron 1p-1h configurations within one even-even nucleus as only those can be populated that can be reached from the ground state of the even-odd target nucleus. Therefore, $(d,p)$ experiments performed along isotopic and isotonic chains are instructive. While these probe neutron configurations above the Fermi surface, $(p,d)$ and $(d,t)$ reactions can be used to study some of the relevant configurations below the Fermi surface.

First $(d,p)$ experiments were performed with the SE-SPS to study the emergence of the PDR around the $N=28$ shell closure. Results for \nuc{62}{Ni} have been published\,\cite{Spi23a}. A complimentary real photon scattering $(\gamma,\gamma^{\prime})$ experiment was performed to aid the identification of the PDR $J^{\pi} = 1^-$ states up to an excitation energy of $E_x = 8.5$\,MeV. As $(d,p)$ data are available up to $S_n$, a follow-up $(\gamma,\gamma^{\prime})$ experiment was performed at the high intensity $\gamma$-ray source (HI$\gamma$S) of the Triangle Universities Nuclear Laboratory (TUNL), which is currently being analyzed. As discussed in \cite{Spi23a}, the combined data allowed us to exclude a significant contribution of the $(2p_{3/2})^{-1}(3s_{1/2})^{+1}$ neutron 1p-1h configuration to the wavefunctions below $S_n$ and, thus, to conclude that any strength increase beyond $N=28$ would need to be linked to either the  $(2p_{3/2})^{-1}(2d_{5/2})^{+1}$ or $(2p_{3/2})^{-1}(2d_{3/2})^{+1}$ configurations if the predictions of Inakura {\it et al.} were correct\,\cite{Ina11a}. 

\begin{figure}[t]
\begin{center}
\includegraphics[width=.95\linewidth]{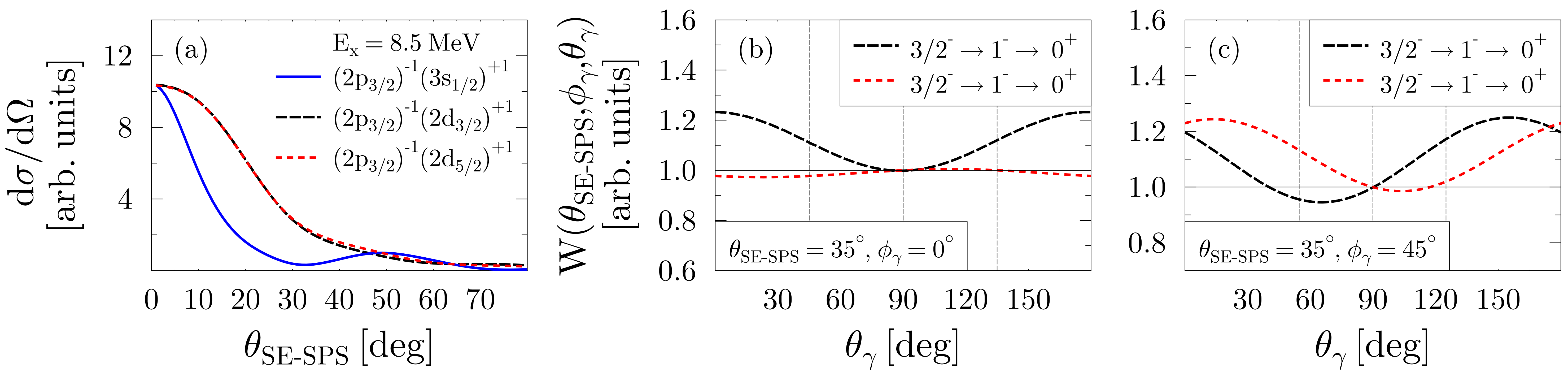}
\end{center}
\caption{(a) Theoretical angular distributions for 1p-1h configurations populating an arbitrary $J^{\pi} = 1^-$ state at $E_x = 8.5$\,MeV in \nuc{53}{Cr}$(d,p)$\nuc{54}{Cr} and calculated with \textsc{chuck3}\,\cite{CHUCK} (b), (c) Predicted proton-$\gamma$ angular correlations in $(d,p\gamma)$ shown for two rings of CeBrA and calculated with \textsc{angcor}\,\cite{angcor}. Some of the detectors in those rings are highlighted with vertical, dashed lines.}\label{fig:pdr}
\end{figure}

While $l$ transfers can be easily determined through $(d,p)$ angular distributions, the $(2p_{3/2})^{-1}(2d_{5/2})^{+1}$ and $(2p_{3/2})^{-1}(2d_{3/2})^{+1}$ neutron 1p-1h configurations cannot be distinguished in SE-SPS singles experiments with an unpolarized deuteron beam (see Fig.\,\ref{fig:pdr}\,(a) for the $(d,p)$ angular distributions calculated with \textsc{chuck3}\,\cite{CHUCK}). Particle-$\gamma$ correlations provide, however, the means to discriminate between spin-orbit partners. See Figs.\,\ref{fig:angcorr}\,(b) and (c) for the particle-$\gamma$ angular correlations calculated with \textsc{angcor}\,\cite{angcor} for a fixed polar angle and two different azimuthal angles. The correlations are expected to look quite different for varying azimuthal angles $\theta_{\gamma}$ and, thus, provide additional sensitivity for discriminating between the spin-orbit partners. As mentioned in Sec.\,\ref{sec:cebra}, the full CeBrA array will enable measurements at different $\theta_{\gamma}$ angles.

$(d,p\gamma)$ experiments have already been performed for nuclei close to $N=28$ with the extended CeBrA demonstrator (see Fig.\,\ref{fig:sesps_cebra}) to study the $\gamma$-ray strength function ($\gamma$SF) via the surrogate reaction method (SRM). As can be seen in Fig.\,\ref{fig:pgmatrix}, the energy resolution of the CeBr$_3$ detectors is sufficient to resolve several low-energy $\gamma$-ray transitions resulting from the deexcitation of low-lying excited states fed by higher-lying states. Therefore, the normalized $\gamma$-ray yields can be determined as a function of excitation energy providing the data for the SRM to constrain the $\gamma$SF\,\cite{Rat19a}. The SE-SPS allows to perform these experiments well past the neutron-separation energy. The indirectly extracted $\gamma$SF from $(d,p\gamma)$ can then be compared to the ground-state $\gamma$SF measured in real-photon scattering, possibly helping to understand whether the PDR is only a feature of the ground state $\gamma$SF. The complimentary $(d,p)$ singles data provide the means to test the microscopic details of wavefunctions predicted by theoretical models that mean to describe the $\gamma$SF as also discussed in \cite{Spi20a, Wei21a}.

\subsection{Nuclear astrophysics studies with CATRiNA}
The CATRiNA neutron detectors are aimed to be used in coincidence with other detector systems at the John D. Fox Laboratory. For instance, we recently performed a resonance spectroscopy study to constrain the $^{25}$Al$(p,\gamma){}^{26}$Si reaction rate via a  very selective $n/\gamma$ coincidence measurement \cite{perello-prc}.

The detection of the long-lived radioisotope $^{26}$Al (5$^+$, T$_{1/2}$ = 7.17$\times$10$^5$ yr) in the Galaxy via the satellite based observation of its characteristic 1.809-MeV $\gamma$-ray line  is of paramount relevance in nuclear astrophysics \cite{prantzos}. This observation is recognized as direct evidence that nucleosynthesis is an ongoing process in the Galaxy, explaining earlier measurements of the excess of $^{26}$Mg found in meteorites and presolar dust grains \cite{lee, jose}. The COMPTEL \cite{comptel} and INTEGRAL \cite{integral} space missions have mapped the intensity distribution of the 1.809-MeV $\gamma$-ray line and inferred an equilibrium mass of $2 - 3$ solar masses of $^{26}$Al in the Milky Way, with most of its mass accumulated in regions of star formation co-rotating with the plane of the Galaxy \cite{diehl}. To understand the stellar nucleosynthesis of $^{26}$Al, one needs to understand all the reactions that produce and destroy $^{26}$Al in the relevant astrophysical scenarios. An additional complication to the accurate modeling and calculation of its nucleosynthesis comes from the short-lived isomeric state in $^{26}$Al (0$^+$, T$_{1/2}$ = 6.4 s) located 228 keV above the long-lived ground state \cite{iliadis}.

At nova burning temperatures of $T \sim 0.1-0.5$\,GK, the $^{25}$Al$(p,\gamma){}^{26}$Si reaction and the subsequent $\beta$-decay of $^{26}$Si leads predominantly to the population of $^{26}$Al in its short-lived isomeric state ($^{26}$Al$^m$) rather than its ground state ($^{26}$Al$^g$). The isomeric $^{26}$Al$^{m}$ (0$^+$) state directly $\beta$-decays to the ground state of $^{26}$Mg (0$^+$), bypassing the emission of the 1.809-MeV $\gamma$-ray line.  Therefore, $^{26}$Al could contribute to the $^{26}$Mg abundance measured in meteorites and pre-solar grains without space telescopes observing its associated $\gamma$ ray.

\begin{figure}
\begin{center}
\includegraphics[width=.95\linewidth]{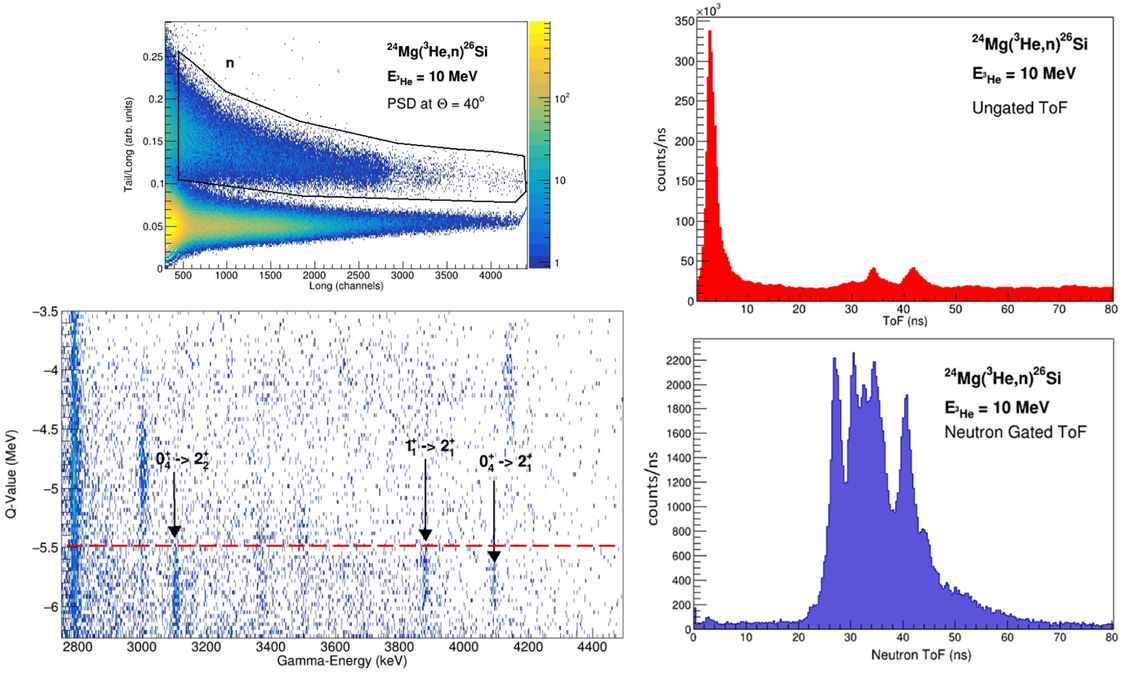}
\end{center}
\caption{A pulse-shape discrimination (PSD) plot for one of the CATRiNA detectors is shown at the top left. Neutron events are clearly separated from $\gamma$-ray events. A raw time-of-flight (ToF) plot of CATRiNA is shown in the top right. 
A ToF plot gated on neutron events is shown in the bottom right. States in $^{26}$Si are identified by their ToF relative to the prompt $\gamma$ ray from the reaction. A zoomed-in portion of a $Q$-value vs $\gamma$-ray energy correlation matrix, built from coincident neutron events and $\gamma$-ray events, is shown in the bottom left. States of interest are below the horizontal, dotted line which indicates the proton-separation energy ($S_p$) in $^{26}$Si. Transitions corresponding to deexcitations of the 0$^{+}_{4}$ and 1$^{+}_{1}$ are clearly visible.}\label{fig:ng-results}
\end{figure}

A high-resolution measurement at the John D. Fox Laboratory was conducted to populate low-lying proton resonances in $^{26}$Si using the $\nuc{24}{Mg}(\nuc{3}{He},n\gamma)\nuc{26}{Si}$ reaction to resolve outstanding discrepancies on the properties of the resonances relevant for the calculation of the $^{25}$Al$(p,\gamma){}^{26}$Si reaction rate. Specifically, we focused on five low-lying resonances within the Gamow window of this reaction \cite{chipps}. For the experiment, a stable 10-MeV $^{3}$He beam from the FN Tandem accelerator was used to bombard an enriched 492-$\mu$g/cm$^{2}$ self-supporting $^{24}$Mg target. The $^{3}$He beam was bunched to 1.7-ns width with intervals of 82.5 ns. The unreacted beam was sent into a thick graphite disk acting as beam-stop located 2\,m downstream from the target position. Neutrons from the $\nuc{24}{Mg}(\nuc{3}{He},n\gamma)\nuc{26}{Si}$ reaction were measured with a set of 16 CATRiNA neutron detectors placed at a distance of 1\,m from the reaction target covering an angular range of $\Delta \theta_{lab}$ = $\pm$40$^{\circ}$. A set of three FSU, Clover-type HPGe $\gamma$-ray detectors, placed at 90$^{\circ}$ from the target, were used to measure $\gamma$ rays from deexcitations of populated states in $^{26}$Si in coincidence. 
The PSD capabilities of CATRiNA were used to separate neutron from $\gamma$ events detected in the CATRiNA detectors. The neutron gate in the PSD plots were then applied to the raw ToF spectra to obtain neutron-ToF spectra for all the CATRiNA detectors as shown in Fig.\,\ref{fig:ng-results}.


The ToF spectrum of each detector cannot be easily added together since neutrons arriving at each detector from a given populated state in $^{26}$Si have different energies due to the reaction kinematics. The neutron events for all 16 CATRiNA detectors were added together in a $Q$-value plot of the reaction. Given that the $Q$-value of the reaction for the ground-state is small ($Q_{gs}$ = 70 keV), the $Q$-value plot can be read as the negative excitation energy of $^{26}$Si. The states of interest, low-lying proton resonances in $^{26}$Si, are located below $Q -5.5$,MeV ($S_P = 5.513$\,MeV). A $Q$-value vs $\gamma$-ray energy correlation matrix was then built for events in coincidence between CATRiNA detectors and the FSU Clover-type HPGe detectors. Several transitions from resonant states are well resolved due to the high resolution of the $\gamma$-ray detectors. An example of this 2D correlation matrix is shown in Fig.\,\ref{fig:ng-results}, expanded on states above the proton-separation threshold (states below the red dotted line) in coincidence with $\gamma$ rays between $2.8 - 4.5$\,MeV. One can clearly identify transitions corresponding to deexcitation of the 0$^{+}_{4}$ and the 1$^{+}_{1}$ states, respectively. Using the extracted spectroscopic information of relevant resonances in $^{26}$Si, we calculated the rate of the $^{25}$Al$(p,\gamma){}^{26}$Si reaction over nova temperatures resolving long standing discrepancies in the literature. See \cite{perello-prc} for more details.

\section{Summary and outlook}

This article highlighted recently commissioned setups for particle-$\gamma$ coincidence experiments at the FSU John D. Fox Superconducting Linear Accelerator Laboratory. Particularly, the combined CeBrA+SE-SPS setup for light-ion transfer experiments and coincident $\gamma$-ray detection, the coupling of the CATRiNA neutron detectors with HPGe detectors measuring neutron-$\gamma$ coincidences for reaction-channel selection, and the combined CLARION2+TRINITY setup for high resolution $\gamma$-ray spectroscopy were featured. These setups allow to perform selective experiments addressing open questions in nuclear structure, nuclear reactions, and nuclear astrophysics. $(d,p)$ studies of single-particle orbitals close to the $N=28$ neutron-shell closure, of the pygmy dipole resonance (PDR), and of the $\nuc{24}{Mg}(\nuc{3}{He},n\gamma)\nuc{26}{Si}$ reaction to resolve outstanding discrepancies on the properties of the resonances relevant for the calculation of the $^{25}$Al$(p,\gamma){}^{26}$Si reaction rate were discussed.  In the next couple of years, the full CeBrA array consisting of 14 CeBr$_3$ detectors will be completed. For the SE-SPS, plans are also in place to design a new focal-plane detector with increased position resolution and higher count rate capabilities based on the multi-layer thick gaseous electron multiplier (M-THGEM) technology \cite{Oli18a, Cor20a, Cir23a}, which also allows for the detection of heavier ions opening new possibilities for experimental studies. In addition, the design of a compact mini-orange conversion electron spectrometer for particle-electron coincidence experiments at the SE-SPS is nearly completed. In the near future, the CATRiNA detectors will be coupled with the CLARION2 HPGe detectors increasing the $\gamma$-ray efficiency significantly compared to previous experiments described in this article. Opportunities for coupling CATRiNA with the SE-SPS for charged-particle-neutron coincidence measurements are also being explored.

\section*{Author Contributions}

M.S. and S.A.C. secured funding for the research presented in this article, supervised research leading to results presented in this article, wrote the manuscript, prepared the figures, and coordinated materials with colleagues at the John D. Fox Laboratory.

\section*{Funding}
The FSU group acknowledges support by the U.S. National Science Foundation (NSF) under Grant Nos. PHY-2012522 [WoU-MMA: Studies of Nuclear Structure and Nuclear Astrophysics] and PHY-2412808 [Studies of Nuclear Structure and Nuclear Astrophysics], and by the U.S. Department of Energy, National Nuclear Security Administration (NNSA) under Grant No. DE-NA0004150 [Center for Excellence in Nuclear Training And University-based Research (CENTAUR)] as part of the Stewardship Science Academic Alliances Centers of Excellence Program. Support by Florida State University is also gratefully acknowledged.

\section*{Acknowledgments}
M.S. and S.A.C. express their gratitude to the staff at the John D. Fox Laboratory for their continued support of the research program. Input from Samuel L. Tabor and Vandana Tripathi for the description of CLARION2+TRINITY, and from John D. Fox Laboratory Director Ingo Wiedenh{\"o}ver is gratefully acknowledged. The authors thank Kirby W. Kemper for input for the history of the laboratory. M.S. thanks Dr. Ben Crider from Mississippi State University for the temporary loan of his $3 \times 3$ inch CeBr$_3$ detectors, and Dr. Lewis A. Riley from Ursinus College as well as Dr. Paul D. Cottle from Florida State University for the continuing and productive collaboration. S.A.C. thanks former FSU graduate students Jesus Perello and Ashton Morelock for the contributions to the development of CATRiNA during their graduate work. M.S. also thanks FSU graduate students Alex L. Conley, Dennis Houlihan, and Bryan Kelly for their contributions to the presented work.

\bibliography{fsu_frontiers}

\end{document}